\documentclass[preprintnumbers,amsmath,amssymb,superscriptaddress]{revtex4}

\usepackage{graphicx}
\usepackage{dcolumn}
\usepackage{bm}
\usepackage{mathrsfs} 

\usepackage[colorlinks=true,linktocpage=true,linkcolor=blue,citecolor=blue]{hyperref}

\begin{document}

\title{The nucleon properties in finite temperature and density with vector meson}

\author{Bingtao Li}
\address{Key Laboratory of Atomic and Subatomic Structure and Quantum Control (MOE), Guangdong Basic Research Center of Excellence for Structure and Fundamental Interactions of Matter, Institute of Quantum Matter, South China Normal University, Guangzhou 510006, China.}
\affiliation{Guangdong-Hong Kong Joint Laboratory of Quantum Matter, Guangdong Provincial Key Laboratory of Nuclear Science, Southern Nuclear Science Computing Center, South China Normal University, Guangzhou 510006, China.}

\author{Yiming Lyu}
\address{Key Laboratory of Atomic and Subatomic Structure and Quantum Control (MOE), Guangdong Basic Research Center of Excellence for Structure and Fundamental Interactions of Matter, Institute of Quantum Matter, South China Normal University, Guangzhou 510006, China.}
\affiliation{Guangdong-Hong Kong Joint Laboratory of Quantum Matter, Guangdong Provincial Key Laboratory of Nuclear Science, Southern Nuclear Science Computing Center, South China Normal University, Guangzhou 510006, China.}

\author{Song Shu}\email{shus@hubu.edu.cn}
\address{School of Physics, Hubei University, Wuhan, Hubei 430062, China.}

\author{PeiXin Weng}
\address{Key Laboratory of Atomic and Subatomic Structure and Quantum Control (MOE), Guangdong Basic Research Center of Excellence for Structure and Fundamental Interactions of Matter, Institute of Quantum Matter, South China Normal University, Guangzhou 510006, China.}
\affiliation{Guangdong-Hong Kong Joint Laboratory of Quantum Matter, Guangdong Provincial Key Laboratory of Nuclear Science, Southern Nuclear Science Computing Center, South China Normal University, Guangzhou 510006, China.}

\author{Hui Zhang}\email{Mr.zhanghui@m.scnu.edu.cn}
\address{Key Laboratory of Atomic and Subatomic Structure and Quantum Control (MOE), Guangdong Basic Research Center of Excellence for Structure and Fundamental Interactions of Matter, Institute of Quantum Matter, South China Normal University, Guangzhou 510006, China.}
\affiliation{Guangdong-Hong Kong Joint Laboratory of Quantum Matter, Guangdong Provincial Key Laboratory of Nuclear Science, Southern Nuclear Science Computing Center, South China Normal University, Guangzhou 510006, China.}
\affiliation{Physics Department and Center for Exploration of Energy and Matter, Indiana University, 2401 N Milo B. Sampson Lane, Bloomington, IN 47408, USA.}

\date{\today}

\begin{abstract}
We introduce the vector meson $\omega$ into the Quark Meson, model and study the impact of vector interactions on the properties of static hadrons using the mean-field approximation. The short-range repulsive force associated with vector interactions leads to an expansion of the root mean square radius of nucleons. While the mass of hadrons increases, the gap between this mass and the energy of the three free constituent quarks decreases, resulting in the instability of hadrons. Our study of nucleon mass and radius at finite temperature and density has potential applications for particle yield in heavy-ion collisions and the mass-radius relationship in compact stars.
\end{abstract}

\maketitle

\section{Introduction}

Quantum Chromodynamics (QCD) is the fundamental theory describing the strong interaction, one of the four fundamental forces in nature. The phase transition of QCD matter refers to the transformation of normal nuclear matter into a quark-gluon plasma (QGP) under conditions of finite temperature and density, accompanied by chiral symmetry restoration and quark deconfinement \cite{Rischke:2003mt, Yagi:2005yb}. Theoretically, due to asymptotic freedom, perturbative QCD calculations are only applicable at short-distance or high-energy scales. Lattice QCD is a significant non-perturbative method; however, it encounters the sign problem in high-density regions. The non-perturbative features of QCD in the low-energy region have not yet led to the establishment of a complete analytical or numerical calculation method \cite{Dong:1995rx, Cohen:1994wm, Meissner:1993ah, Alkofer:2000wg, Bhagwat:2007ha}. These challenges are closely related to the non-perturbative structure of the QCD vacuum. Consequently, effective models that incorporate the non-perturbative properties of low-energy QCD have become important approaches, including the linear Sigma model (LSM) \cite{Gell-Mann:1960mvl}, the Nambu-Jona-Lasinio (NJL) model \cite{Nambu:1961tp}, the Brueckner-Hartree-Fock (BHF) theory \cite{Bethe:1956zz}, and the relativistic mean-field (RMF) model \cite{Serot:1984ey, Serot:1997xg}, and so on.

In recent years, the study of hadronic properties through heavy-ion collisions, such as those conducted at the Relativistic Heavy Ion Collider (RHIC) at Brookhaven National Laboratory and the Large Hadron Collider (LHC) at CERN, has gained prominence. Additionally, astronomical observations of compact stars have contributed to constraining the equation of state of hadronic matter. The hot and dense vacuum produced by relativistic heavy-ion collisions and neutron stars differs significantly from the vacuum at zero temperature and density \cite{Gyulassy:2004zy, Kapusta:2007qb}. The nuclear interactions at finite density are highly complex, with high densities causing hadron wave functions to overlap, making repulsive interactions increasingly significant~\cite{Fukushima:2020cmk}. Thus, vector interactions play a crucial role in studying hadronic properties.

The vector interaction has been incorporated into various effective models to study the properties of hadrons.\cite{Nagai:2006ya} The vector mesons $\omega$ and $\rho$ have been integrated into the Quark Meson model to study the phase transition of QCD using the functional renormalization group method. \cite{Zhang:2017icm,CamaraPereira:2020xla} The quark mass density-dependent (QMDD) model with the $\omega$ meson has been explored in Ref.~\cite{Wu:2007je}. The linear Sigma model (LSM), also known as the chiral soliton model, is characterized by its ability to describe the chiral properties of the vacuum and its spontaneous symmetry breaking. In the mean-field approximation, this model has a semiclassical soliton solution, corresponding to the chiral soliton, from which the static properties of hadrons can be derived \cite{Birse:1984loi, Cohen:1986va, Goeke:1988hp, Aly:1998wg, Broniowski:2002ew,Mao:2013qu,Abu-Shady:2012ewe}. In the original LSM, only scalar mesons $\sigma$ and pseudo-scalar mesons $\pi$ are included, where quarks provide long-range attraction through the exchange of scalar mesons, neglecting the short-range repulsion between quarks. 

In this work, we introduce the vector meson $\omega$ into the quark meson model, and examine the impact of vector interactions (primarily providing repulsive forces) on nucleon properties. Our findings reveal that both the nucleon mass and radius increase with the introduction of vector coupling. The structure of this work is as follows. In Section~\ref{sec:model}, we introduce the chiral soliton model with vector interactions at zero temperature and zero density, extending it to finite temperature and density in Section~\ref{sec:finite}. In Section~\ref{sec:result}, we present the thermal effective potential density and display the soliton solutions of the chiral soliton equation under different conditions. The properties of the static hadrons are then discussed before the conclusions.

\section{Model}  \label{sec:model}

In Minkowski space, the chiral Lagrangian of an effective Quark Meson model with vector interactions is given by \cite{Wu:2007je, Zhang:2017icm}:
\begin{equation}
    \mathcal{L}=\Bar{\psi}[i\gamma^\mu \partial_\mu +g (\hat{\sigma}+i\gamma
    _5 \Vec{\tau} \cdot\Vec{\hat{\pi}})-g_\omega \gamma_\mu \hat{\omega}^\mu]\psi+\frac{1}{2}(\partial_\mu \hat{\sigma}   \partial^\mu    \hat{\sigma}  +\partial_\mu \Vec{\hat{\pi}} \partial^\mu \Vec{\hat{\pi}})-\frac{1}{4}F_{\mu\nu} F^{\mu\nu}- U(\hat{\sigma},\Vec{\hat{\pi}},\hat{\omega}) . \label{eq:1}
\end{equation}
Here, the Lagrangian is invariant under chiral $SU(2)_R \times SU(2)_L$ symmetry transformations if the explicit symmetry breaking term $H \sigma$ is zero, and the field strength tensor $F_{\mu\nu} = \partial_\mu \hat{\omega}_\nu - \partial_\nu \hat{\omega}_\mu$ represents the gauge field corresponding to the interactions. The quark field $\psi$ has spin $\frac{1}{2}$ and two flavors, $\psi = (u, d)^\top$. The field $\hat{\sigma}$ represents the isospin singlet scalar meson field with spin $0$, while $\vec{\hat{\pi}}$ is the isospin vector meson field with spin $0$, $\vec{\hat{\pi}} = (\hat{\pi}_1, \hat{\pi}_2, \hat{\pi}_3)$. The vector field $\hat{\omega}$ has spin $1$. The potential function is given by:
\begin{equation}
        U(\hat{\sigma},\Vec{\hat{\pi}},\hat{\omega})=\frac{\lambda}{4}(\hat{\sigma}^2+\Vec{\hat{\pi}}^2-v^2)+H\hat{\sigma}-\frac{1}{2}m^2_\omega \hat{\omega}_\mu \hat{\omega}^\mu . \label{eq:2}
\end{equation}
In this formulation, $f_\pi = 93\ \rm{MeV}$ is the decay constant of the $\pi$ meson, and $H = f_\pi m^2_\pi$ represents the explicit chiral symmetry breaking term, with $m_\pi = 138\ \rm{MeV}$ being the mass of the $\pi$ meson. The chiral symmetry of the vacuum is explicitly broken. Due to rotational symmetry, the $\omega$ meson has only the zero component $\omega = \omega^0$~\cite{Floerchinger:2012xd}. The vacuum expectation values of the mesons are $\langle\hat{\sigma}\rangle = -f_\pi$, $\langle\vec{\hat{\pi}}\rangle = 0$, and $\langle \hat{\omega} \rangle = 0$. 

In the vacuum, the constituent quark mass is $M_q = gf_\pi$, and the mass of the $\sigma$ meson is defined as $m^2_\sigma = m^2_\pi + 2\lambda f^2_\pi$. The parameter $v^2$ is given by $v^2 = f^2_\pi - \frac{m^2_\pi}{\lambda}$.We treat $g_\omega$ and $m_\omega$ as effective quantities, typical values are $m_\omega \approx 1\ \rm{GeV}$, and $g_\omega$ ranging from $1$ to $10$, resulting in $g_\omega / m_\omega \approx 10^{-3} - 10^{-2}\ \rm{MeV}^{-1}$. In our calculations, we adopt parameters from Birse's work \cite{Birse:1984loi}, setting the constituent quark mass to $500\ \rm{MeV}$, the $\sigma$ meson mass to $m_\sigma = 1200\ \rm{MeV}$, yielding $g \approx 5.28$ and $\lambda \approx 82.1$.

Starting from the Lagrangian, we derive the radial equations of motion for the quark and meson fields \cite{Birse:1984loi}:
\begin{eqnarray}
    \frac{du(r)}{dr} = -(\epsilon - g\sigma(r))v(r) - g\pi(r)u(r) + g_\omega \omega(r)v(r)&, \label{eq:ur} \\
    \frac{dv}{dr} = -(\frac{2}{r} - g\pi(r))v(r) + (\epsilon + g\sigma(r))u(r) - g_\omega \omega(r)u(r)&, \label{eq:vr} \\
    \frac{d^2\sigma(r)}{dr^2} + \frac{2}{r}\frac{d\sigma(r)}{dr} + Ng(u^2(r) - v^2(r)) = \frac{\partial U}{\partial \sigma}&, \label{eq:sigmar} \\
    \frac{d^2\pi(r)}{dr^2} + \frac{2}{r}\frac{d\pi(r)}{dr} - \frac{2\pi(r)}{r^2} + 2Ngu(r)v(r) = \frac{\partial U}{\partial \pi}& , \label{eq:pir} \\
    \frac{d^2\omega(r)}{dr^2} + \frac{2}{r}\frac{d\omega(r)}{dr} + Ng(u^2(r) + v^2(r)) = \frac{\partial U}{\partial \omega}&. \label{eq:omegar}
\end{eqnarray}
In these equations, we employ the mean-field approximation and the "hedgehog" ansatz:
\begin{eqnarray}
    \langle\hat{\sigma}(\Vec{r},t)\rangle=\sigma(r),\quad \langle\Vec{\hat{\pi(r,t)}}\rangle=\vec{\hat{r}}\pi(r),\quad
    \langle\hat{\omega}(\Vec{r},t)\rangle=\omega(r) ,  \\
    \psi(\Vec{r},t)=e^{-i\epsilon t}\sum^N_{i=1}q_i(\Vec{r}), \quad q_i(\Vec{r})=\begin{pmatrix} u(r)   \\
    i\Vec{\sigma} \cdot \vec{\hat{r}}v(r)    \end{pmatrix} \chi ,   \\
    (\Vec{\sigma}+\Vec{\tau})\chi=0 .
\end{eqnarray}
Here, $q_i$ represents $N$ equivalent lowest-energy $s$-wave constituent quarks with an eigenenergy $\epsilon$. For baryons and mesons, we set $N$ to $3$ and $2$, respectively. $\chi$ denotes the spinor. The wave function of the quarks satisfies the normalization condition:
\begin{equation}
        4\pi\int r^2(u^2(r)+v^2(r))dr=1 .
        \label{eq:normalization}
\end{equation}

The equations of motion for the radial direction of the quark and meson fields satisfy the following boundary conditions:
\begin{eqnarray}
    v(0)=0,\quad \left. \frac{d \sigma(r)}{d r} \right|_{r=0} =0 , \quad \pi(0)=0, \quad \left. \frac{d \omega(r)}{d r} \right|_{r=0} =0, \nonumber \\
        u(\infty)=0,\quad \sigma(\infty)=-f_\pi,\quad \pi(\infty)=0, \quad \omega(\infty)=0 .
        \label{eq:boundary} 
\end{eqnarray}

The asymptotic vacuum expectation values in the soliton field boundary conditions are determined by the vacuum at infinity, where the physical vacuum is chirally broken. The equations for the quark and meson fields, along with the normalization condition and boundary conditions for the quark field, together constitute a highly nonlinear set of coupled equations that can be solved numerically.

\section{Chiral Soliton at finite temperature and density} \label{sec:finite}

In order to investigate the effects of temperature and density on chiral solitons, we embed a soliton into a uniform hot dense quark medium with temperature of $T$ and chemical potential of $\mu$. We first derive the thermodynamic potential of the uniform hot background at finite temperature and density using the finite temperature field theory \cite{Kapusta:2007qb}. The grand canonical partition function of the system is expressed as:
\begin{eqnarray}
    \mathcal{Z}&=\rm{Tr}\exp{[-\hat{\mathcal{H}}-\mu \hat{\mathcal{N}}/T]} =\int \prod \limits_{j} \mathcal{D}\sigma \mathcal{D}\omega \mathcal{D}\pi_j \int \mathcal{D}\psi \mathcal{D}\Bar{\psi} \exp{[\int_{x} \mathcal{L}+\mu \Bar{\psi}\gamma^0\psi)]} ,
    \label{eq:13}
\end{eqnarray}
where $j=1,2,3$, $\int_x=i\int_{0}^{1/T} dt \int_V d^3x$ and $V$ is the volume of the system.

Under the mean-field approximation, the vacuum expectation values of the meson fields $\sigma$, $\pi$, and $\omega$, which do not depend on time, are treated as classical mean fields, neglecting quantum and thermal fluctuations. The thermodynamic potential is then given by:
\begin{equation}
    \Omega(\sigma,\boldsymbol{\pi},\omega,T,\mu)=\frac{-T\ln{\mathcal{Z}}}{V}=U(\sigma,\boldsymbol{\pi},\omega)+\Omega_{\Bar{\psi} \psi} ,
    \label{eq:potential}
\end{equation}
where the contribution of (anti)quarks at finite temperature and density is:
\begin{equation}
    \Omega_{\Bar{\psi} \psi}=-\nu_q T \int \frac{d^3\boldsymbol{p}}{(2\pi)^3}\{\ln{[1+e^{-(E_q-\mu_{eff})/T}]}+\ln{[1+e^{-(E_q+\mu_{eff})/T}]}\} ,
    \label{eq:quarks}
\end{equation}
where $\nu_q$ is the degeneracy factor, given by $\nu_q = 2(spin) \times 2 (flavor) \times 3 (color) = 12$. $E_q = \sqrt{\vec{p}^2 + M^2_q}$ represents the energy of the valence (anti)quarks $u$ and $d$. The constituent quark mass $M_q^2 = g^2 (\sigma^2+\vec{\pi}^2)$, where $\sigma_v$ is the expectation value of the sigma field. The effective chemical potential is defined as \cite{Skokov:2010sf}:
\begin{equation}
    \mu_{eff}=\mu-g_\omega\omega.
    \label{eq:MuEff}
\end{equation}

The thermodynamic vacuum at finite temperature and density occurs at $\sigma = \sigma_v$, $\pi = 0$, and $\omega = \omega_v$, where these values are determined by the saddle point of the thermodynamic potential, equivalently the gap equations:
\begin{equation}
    \frac{\partial \Omega}{\partial \sigma} = \frac{\partial \Omega}{\partial \pi} = \frac{\partial \Omega}{\partial \omega} = 0 .
    \label{eq:gap}
\end{equation}

Next, we embed a soliton into the uniform hot dense background, by simply replacing $U(\sigma,\boldsymbol{\pi}, \omega)$ with $\Omega(\sigma, \boldsymbol{\pi}, \omega, T, \mu)$. This leads to a new set of equations of motion for the chiral soliton:
\begin{eqnarray}
    \frac{d^2\sigma(r)}{dr^2}+\frac{2}{r}\frac{d\sigma(r)}{dr}+Ng(u^2(r)-v^2(r))=\frac{\partial \Omega}{\partial \sigma} , \nonumber \\
    \frac{d^2\pi(r)}{dr^2}+\frac{2}{r}\frac{d\pi(r)}{dr}-\frac{2\pi(r)}{r^2}+2Ngu(r)v(r)=\frac{\partial \Omega}{\partial \pi} , \nonumber \\
    \frac{d^2\omega(r)}{dr^2}+\frac{2}{r}\frac{d\omega(r)}{dr}+Ng(u^2(r)+v^2(r))=-\frac{\partial \Omega}{\partial \omega} ,
    \label{eq:motion2}
\end{eqnarray}
where
\begin{eqnarray}
        \frac{\partial \Omega}{\partial \sigma}=\frac{\partial U}{\partial \sigma}+g\rho_s , \\
       \frac{\partial \Omega}{\partial \pi}=\frac{\partial U}{\partial \pi}+g\rho_{ps} , \\
        \frac{\partial \Omega}{\partial \omega}=\frac{\partial U}{\partial \omega}+g_\omega\rho .
        \label{eq:Omega}
\end{eqnarray}
The density of scalar, pseudoscalar, and vector current of (anti)quarks are expressed as:
\begin{eqnarray}
    \rho_s=\langle \bar{q} q\rangle=g\sigma v_q \int \frac{d^3 \boldsymbol{p}}{(2\pi)^3} \frac{1}{E_q}
        \left(\frac{1}{1+e^{(E_q-\mu_{eff}/T)}}+\frac{1}{1+e^{(E_q+\mu_{eff}/T)}}\right) , \\
    \rho_{ps}=\langle \bar{q} i\gamma_5 \boldsymbol{\tau} q\rangle=g\boldsymbol{\pi} v_q \int \frac{d^3 \boldsymbol{p}}{(2\pi)^3} \frac{1}{E_q}
        \left(\frac{1}{1+e^{(E_q-\mu_{eff}/T)}}+\frac{1}{1+(e^{(E_q+\mu_{eff})/T)}}\right) , \\
    \rho=\langle \bar{q} \gamma^0 q\rangle=g_\omega  v_q \int \frac{d^3 \boldsymbol{p}}{(2\pi)^3} \frac{1}{E_q}
        \left(\frac{1}{1+e^{(E_q-\mu_{eff}/T)}}-\frac{1}{1+(e^{(E_q+\mu_{eff})/T)}}\right) .
\end{eqnarray}

Boundary conditions become:
\begin{eqnarray}
    v(0)=0,\quad \left. \frac{d \sigma(r)}{d r} \right|_{r=0} =0 , \quad \pi(0)=0, \quad \left. \frac{d \omega(r)}{d r} \right|_{r=0} =0, \\
    u(\infty)=0,\quad \sigma(\infty)=\sigma_v,\quad \pi(\infty)=0, \quad \omega(\infty)=\omega_v .
    \label{eq:33}
\end{eqnarray}

By simultaneously solving the quark field Eqs.~(\ref{eq:ur}), (\ref{eq:vr}) and the meson field Eqs.~(\ref{eq:motion2}), along with the normalization and boundary conditions, we can obtain numerical solutions for soliton solutions with vector interactions.

In this statement, we present the physical picture of the system. At zero temperature and zero density, a soliton with vector interactions is embedded in a pure physical vacuum described by the potential energy $U(\sigma, \boldsymbol{\pi}, \omega)$. The physical vacuum is determined by the gap Eq.~(\ref{eq:gap}). At large radii, the fields $\sigma$, $\pi$, and $\omega$ exhibit their vacuum expectation values, which vary with the radius inside the soliton. Outside the soliton, the potential energy consists of a constant field composed of the vacuum expectation values of the mesons. At finite temperature and density, the effective potential energy $U(\sigma, \boldsymbol{\pi}, \omega)$ is replaced by the thermal effective potential $\Omega(\sigma, \boldsymbol{\pi}, \omega, T, \mu)$. Inside the soliton, the fields $\sigma$, $\pi$, and $\omega$ vary with radius, while outside the soliton, there exists a thermal average field background, with the meson fields exhibiting their thermal physical vacuum expectation values.

By solving the soliton solutions with three-quark vector interactions and minimizing the energy, we can determine the total energy $E_B$ of the system in the thermal background. The thermal potential is shifted to zero by subtracting a constant $\Omega(\sigma_v,\pi_v,\omega_v,T,\mu)$ to derive a finite and well-defined baryon mass, ensuring that the pressure at the surface of the solitons in both the vacuum and the thermal medium remains zero \cite{Zhang:2014wta,Zhang:2015vva}.
\begin{equation}
    E_B=N\epsilon+4\pi \int dr r^2[\frac{1}{2} \left(\frac{d\sigma}{dr}\right)^2+\frac{1}{2} \left(\frac{d\pi}{dr}\right)^2+\frac{\pi^2}{r^2}-\left(\frac{d\omega}{dr}\right)^2+\Omega(\sigma(r),\pi(r),\omega(r),T,\mu) -\Omega(\sigma_v, \pi_v , \omega_v , T, \mu)] .
    \label{eq:27}
\end{equation}
The root mean square (RMS) radius of the baryon, $R = \sqrt{\langle r^2 \rangle}$:
\begin{equation}
    \langle r^2 \rangle=4\pi \int_0^\infty r^4(u^2+v^2)dr .
    \label{eq:28}
\end{equation}

\section{Results}\label{sec:result}

\begin{figure}[htb]
    \centering
    \includegraphics[width=0.5\textwidth]{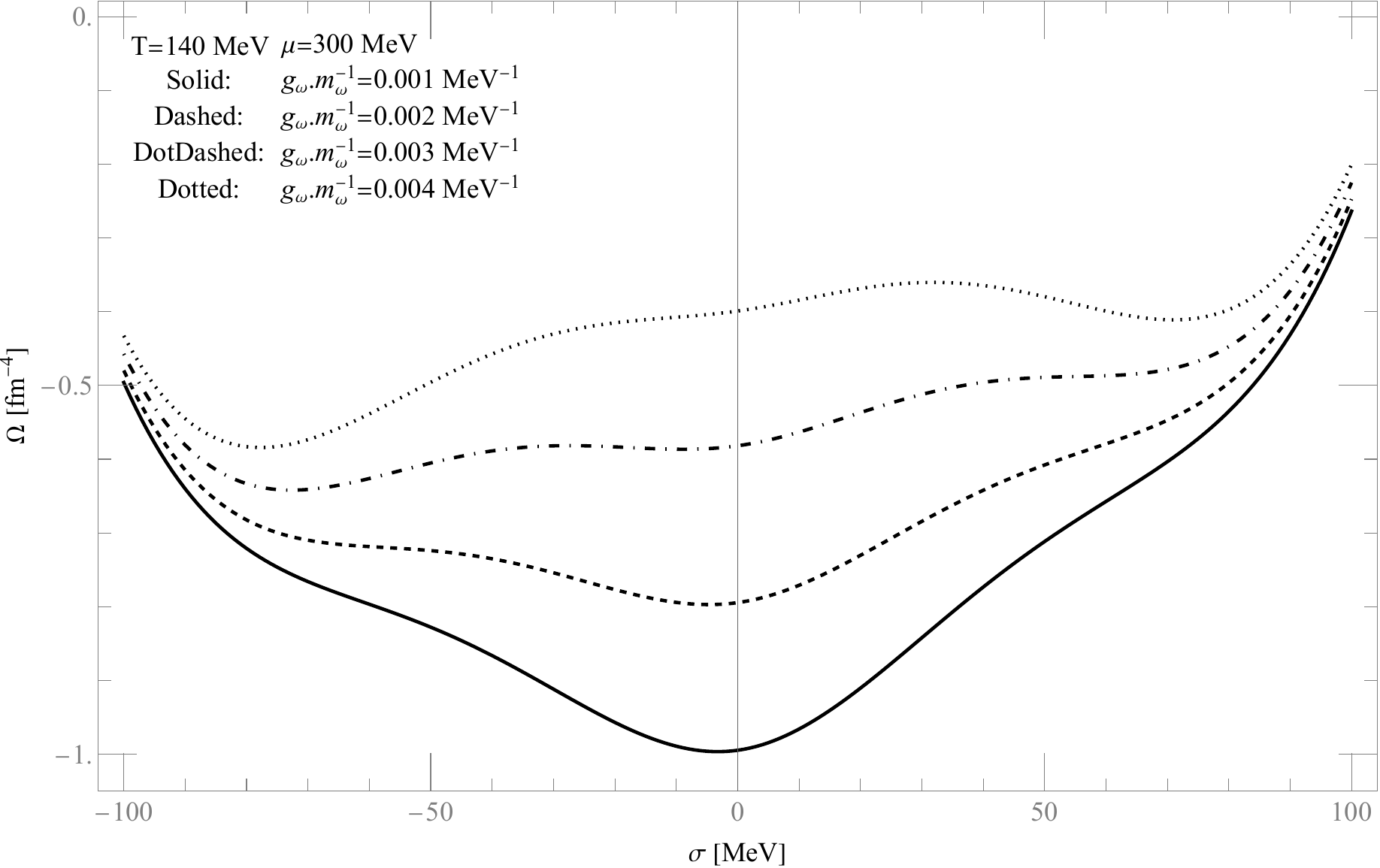}
    \caption{Thermal effective potential $\Omega$ as a function of the order parameter $\sigma$ at $T=140\ \rm{MeV},\ \mu=300\ \rm{MeV}$, and $g_\omega/m_\omega=\{1,\ 2,\ 3,\ 4\} \times 10^{-3} \ \rm{MeV}^{-1}$. } 
    \label{fig1}
\end{figure}
In Fig.~\ref{fig1}, we plot the thermal effective potential $\Omega$ as a function of the order parameter $\sigma$ at $T=140\ \rm{MeV},\ \mu=300\ \rm{MeV}$, and $g_\omega/m_\omega=\{1,\ 2,\ 3,\ 4\} \times 10^{-3} \ \rm{MeV}^{-1}$. As the vector coupling constant increases, the phase transition is postponed. When $T<T_c$ and $\mu<\mu_c$, the thermal effective potential exhibits a single absolute minimum, corresponding to the chiral symmetry broken (CSB) vacuum. In this model with the mean-field approximation method, the transition from the broken chiral vacuum to the restoration of chiral symmetry is classified as a first-order phase transition.

\begin{figure}[htb]
    \centering
    \includegraphics[width=0.45\textwidth]{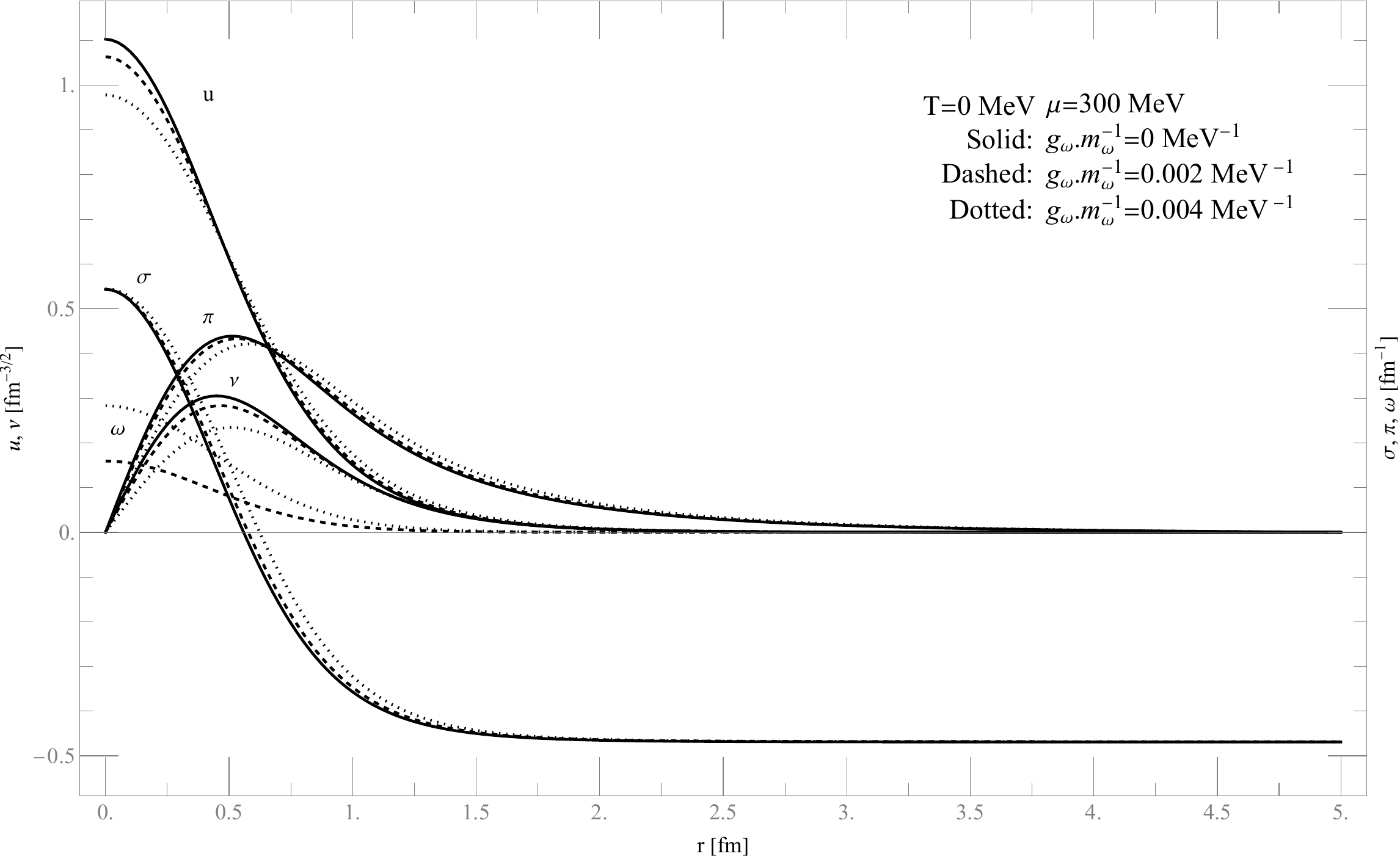}
    \includegraphics[width=0.45\textwidth]{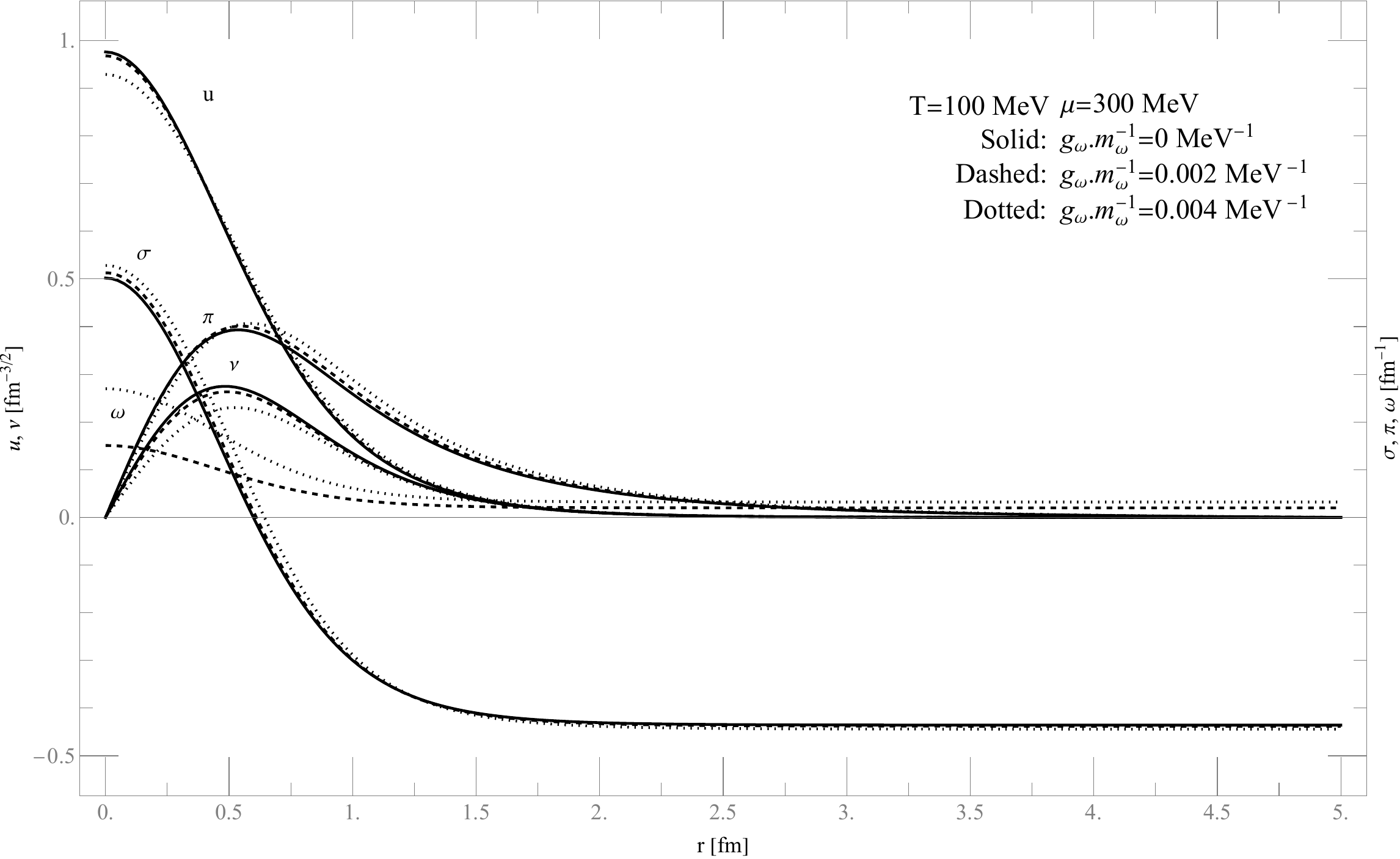}
    \caption{The quark fields $u(r)$, $v(r)$, and the meson fields $\sigma(r)$, $\pi(r)$, $\omega(r)$ as functions of $r$ for $T=0\ \rm{MeV}$ (left), and $T=100\ \rm{MeV}$ (right) with a fixed chemical potential $\mu=300\ \rm{MeV}$ at $g_\omega/m_\omega=\{0,\ 2,\ 4\} \times 10^{-3}\ \rm{MeV}^{-1}$. }
    \label{fig2}
\end{figure}

From this analysis, it is evident that a hadronic phase exists when the CSB vacuum is the physical vacuum. We present soliton solutions with vector interaction at finite temperatures and densities in Fig.~\ref{fig2}. The quark fields $u(r)$, $v(r)$, and the meson fields $\sigma(r)$, $\pi(r)$, $\omega(r)$ as functions of $r$ for $T=0\ \rm{MeV}$ (left), and $T=100\ \rm{MeV}$ (right) with a fixed chemical potential $\mu=300\ \rm{MeV}$ at $g_\omega/m_\omega=\{0,\ 2,\ 4\} \times 10^{-3}\ \rm{MeV}^{-1}$ are plotted. In the CSB case, as the radius approaches infinity, the values of $\sigma(r)$ and $\omega(r)$ in the soliton solutions converge to the physical vacuum values. As the vector coupling constant increases, all wave functions extend to larger $r$, indicating an increase in the radius, which will be discussed further below. Moreover, the increase in $\omega$ mesons is greater than that of $\sigma$ mesons, suggesting that at high temperatures and densities, the contribution of vector mesons becomes significant and cannot be neglected.

\begin{figure}[htb]
    \centering
    \includegraphics[width=0.45\textwidth]{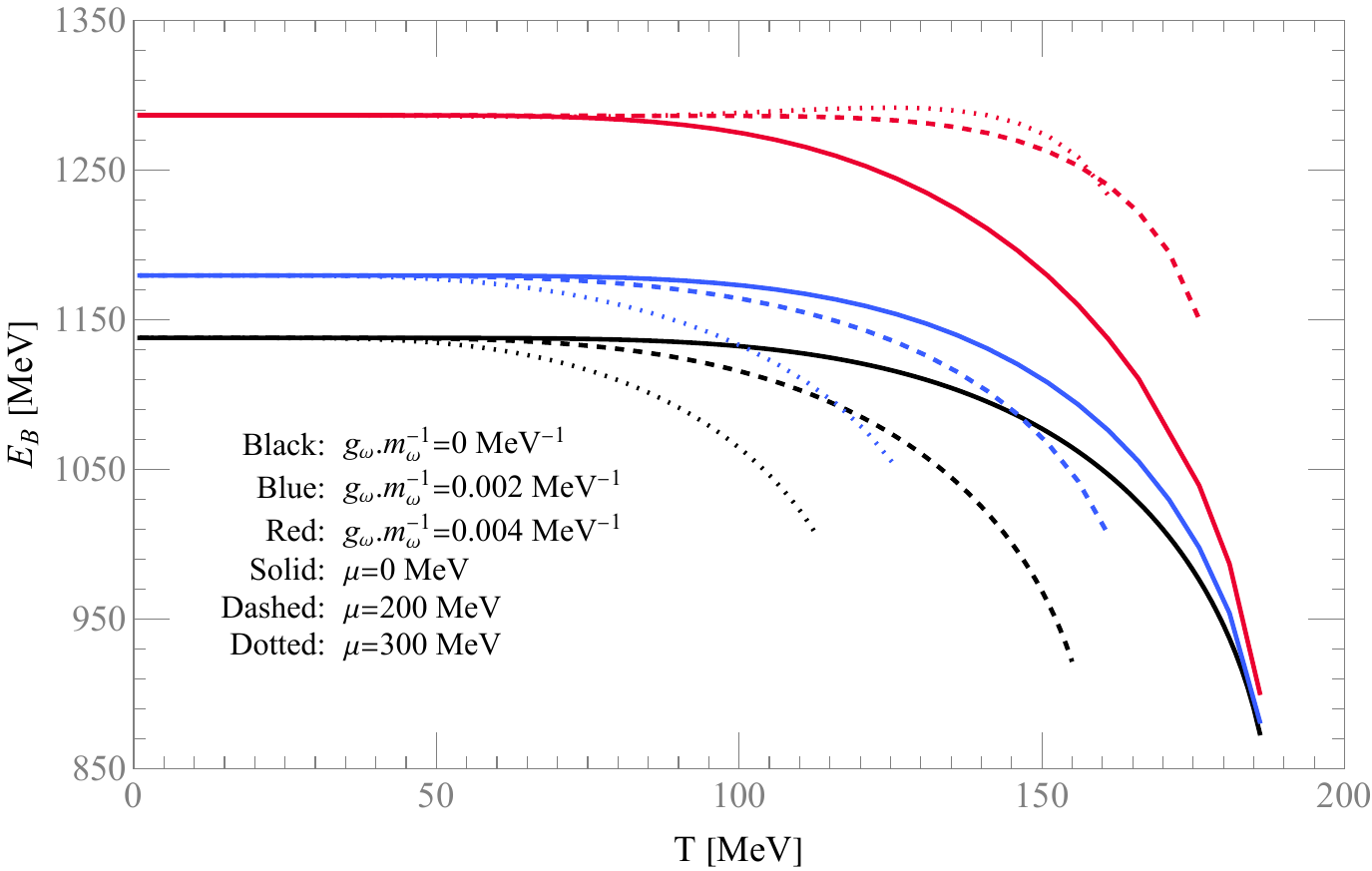}
    \includegraphics[width=0.45\textwidth]{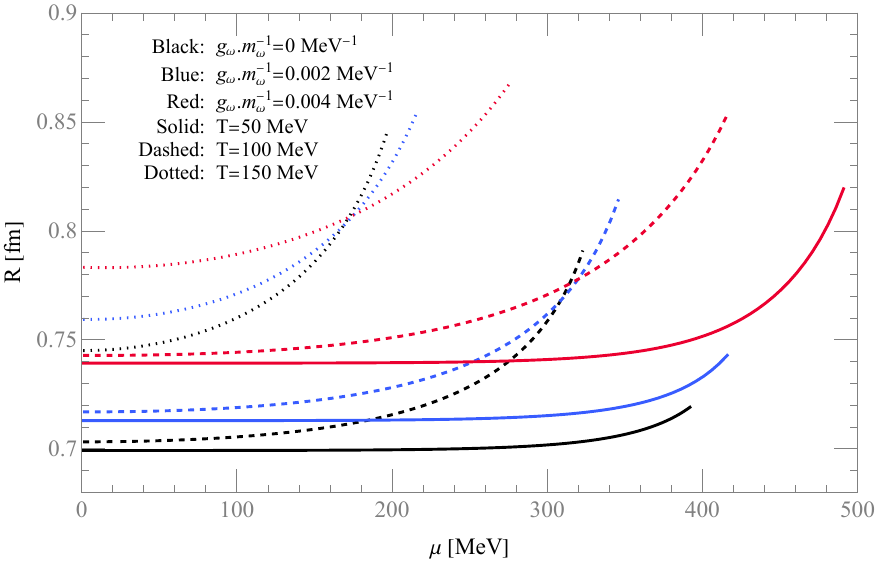}
    \caption{The RMS radius $R$ for $g_\omega/m_\omega=\{0,\ 2,\ 4\} \times 10^{-3}\ \ \rm{MeV}^{-1}$: (left) as a function of temperature $T$ with chemical potential fixed at $\mu=\{0,\ 200,\ 300\}\ \rm{MeV}$, (right) as a function of chemical potential $\mu$ with temperature fixed at $T=\{50,\ 100,\ 150\}\ \rm{MeV}$. }
    \label{fig3}
\end{figure}
\begin{figure}[htb]
    \centering
    \includegraphics[width=0.45\textwidth]{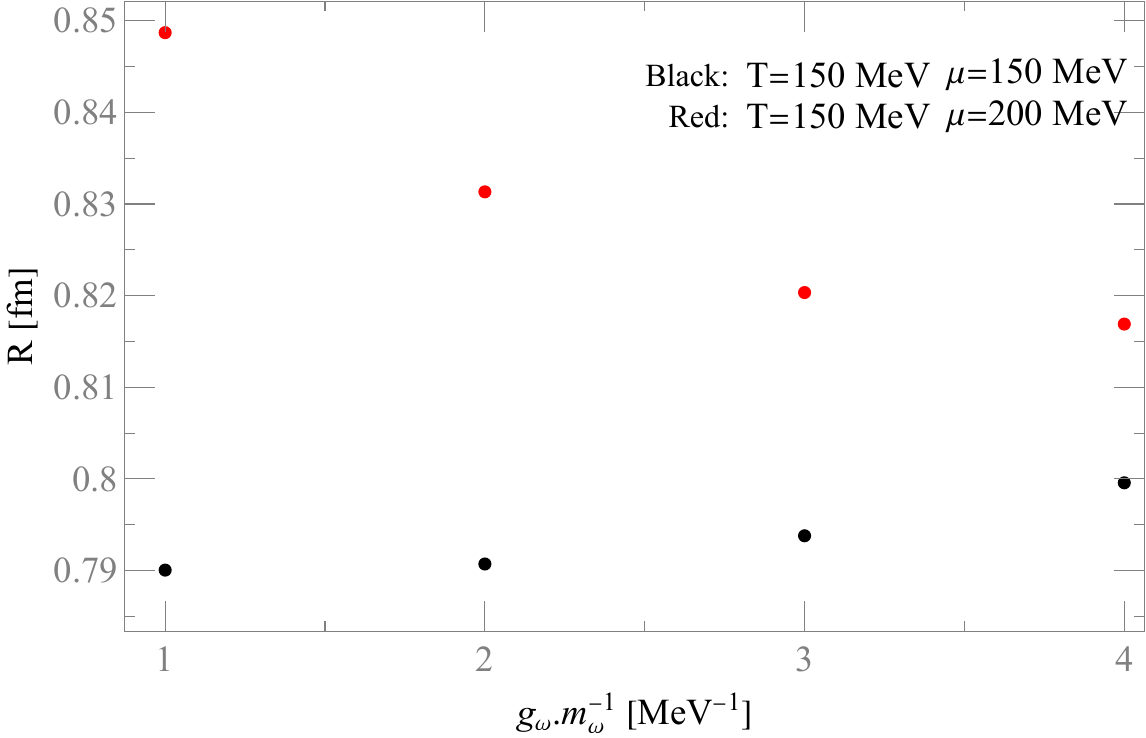}
    \caption{The RMS radius $R$ as a function of $g_\omega/m_\omega$, with temperature and chemical potential fixed at $T =150\ \rm{MeV},\ \mu=\{150,\ 200\} \ \rm{MeV}$.}
    \label{fig5}
\end{figure}

In Fig. \ref{fig3}, we plot the RMS radius $R$ for $g_\omega/m_\omega=\{0,\ 2,\ 4\} \times 10^{-3}\ \ \rm{MeV}^{-1}$: (left) as a function of temperature $T$ with chemical potential fixed at $\mu=\{0,\ 200,\ 300\}\ \rm{MeV}$, (right) as a function of chemical potential $\mu$ with temperature fixed at $T=\{50,\ 100,\ 150\}\ \rm{MeV}$. This provides insight into the expansion of nucleons as temperature and density increase. We observe intersection points in these curves, indicating that the radius increases with vector couplings at low temperatures (chemical potential) for fixed chemical potential (temperature), while it decreases with vector couplings at high temperatures (chemical potential) for fixed chemical potential (temperature). Consequently, we also plot the RMS radius $R$ as a function of $g_\omega/m_\omega$, with temperature and chemical potential fixed at $T = 150\ \rm{MeV},\ \mu=\{150,\ 200\}\ \rm{MeV}$ in Fig.~\ref{fig5}. This unusual behavior can be explained by competitive effects. For the case of $T =150\ \rm{MeV},\ \mu=150\ \rm{MeV}$, the short-range repulsive force between quarks, supported by vector fields (specifically the $\omega$ meson in this study), dominates. Conversely, for the case of $T =150\ \rm{MeV},\ \mu=200\ \rm{MeV}$, the long-range attractive force, supported by scalar fields (the $\sigma,\ \pi$ mesons), becomes dominant.

\begin{figure}[htb]
    \centering
    \includegraphics[width=0.5\textwidth]{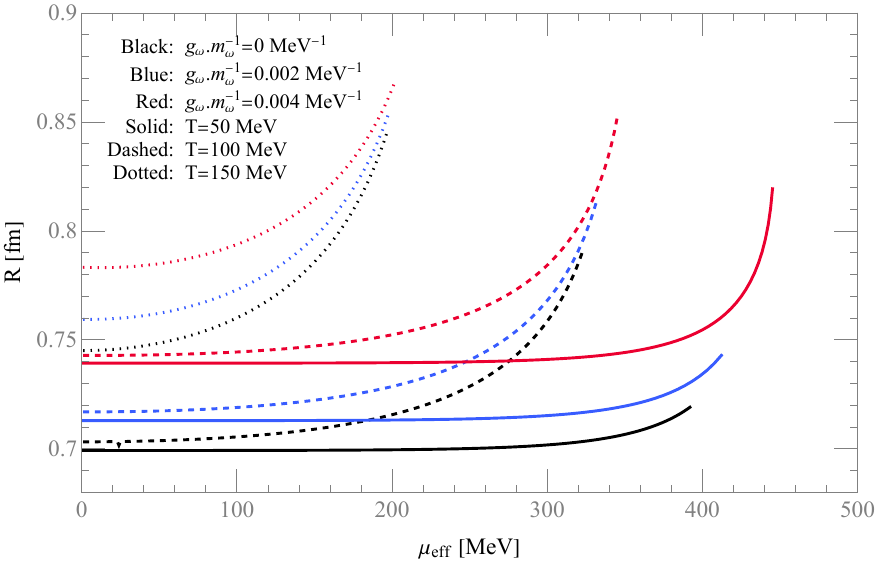}
    \caption{The RMS radius $R$ as a function of effective chemical potential $\mu_{eff}$, with temperature fixed at $T=\{50, 100, 150\}\ \rm{MeV}$ and $g_\omega/m_\omega=\{0, 2, 4\} \times 10^{-3}\ \rm{MeV}^{-1}$. }
    \label{fig6}
\end{figure}

To isolate the contribution of scalar fields, we examine the RMS radius $R$ as a function of effective chemical potential with temperature fixed at $T=\{50, 100, 150\}\ \rm{MeV}$ and $g_\omega/m_\omega=\{0, 2, 4\} \times 10^{-3}\ \rm{MeV}^{-1}$ in Fig.~\ref{fig6}. The effective chemical potential $\mu_{eff}$ is defined in Eq.~(\ref{eq:MuEff}). The repulsive force provided by vector interactions does lead to an increase in the nucleon RMS radius.

\begin{figure}[htb]
    \centering
    \includegraphics[width=0.45\textwidth]{EB_T.pdf}
    \includegraphics[width=0.45\textwidth]{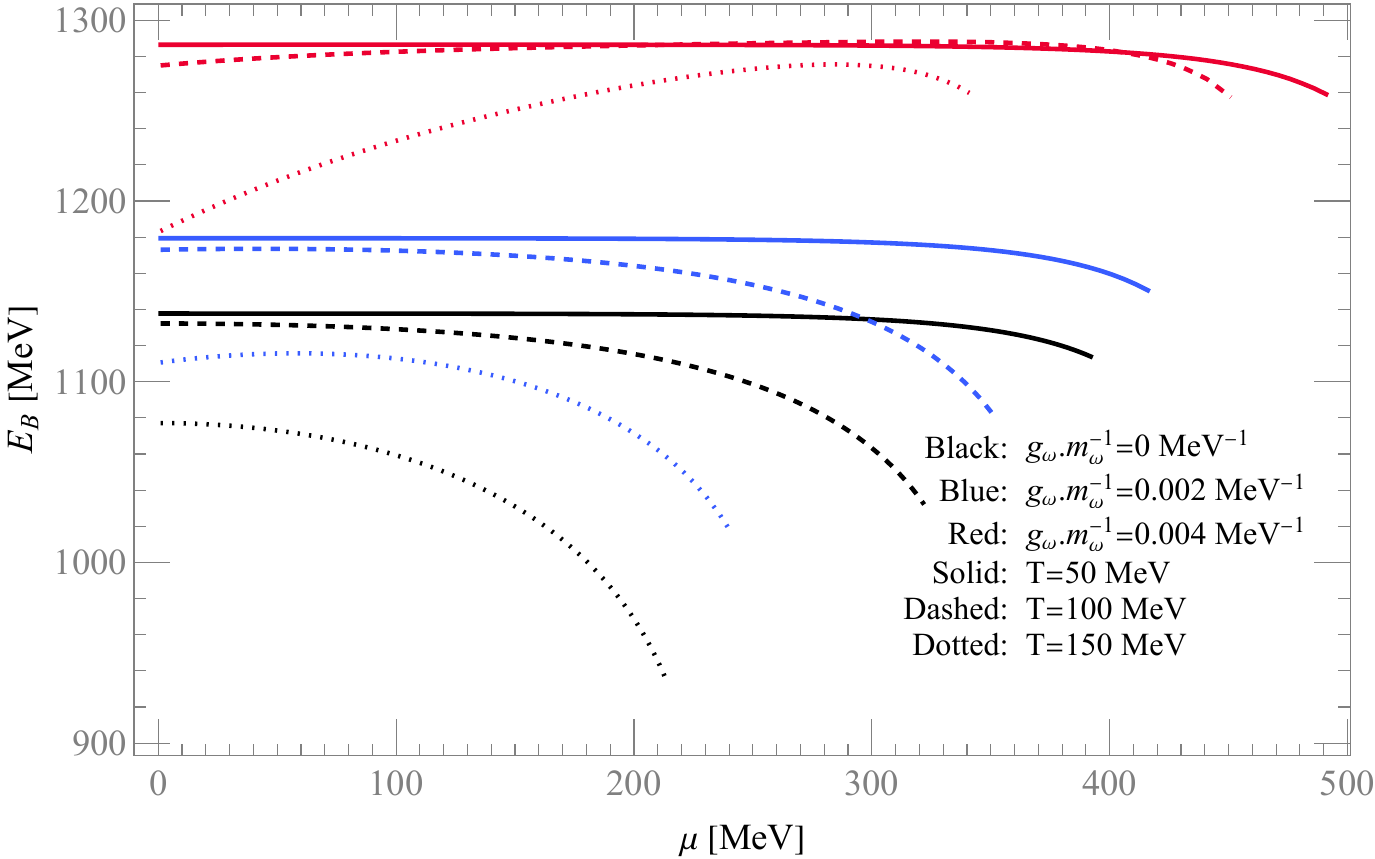}
    \caption{The nucleon mass $E_B$ for $g_\omega/m_\omega=\{0,\ 2,\ 4\} \times 10^{-3}\ \ \rm{MeV}^{-1}$: (left) as a function of temperature $T$ with chemical potential fixed at $\mu=\{0,\ 200,\ 300\}\ \rm{MeV}$, (right) as a function of chemical potential $\mu$ with temperature fixed at $T=\{50,\ 100,\ 150\}\ \rm{MeV}$. }
    \label{fig4}
\end{figure}
In Fig. \ref{fig4}, we plot the nucleon mass $E_B$ for $g_\omega/m_\omega=\{0,\ 2,\ 4\} \times 10^{-3}\ \ \rm{MeV}^{-1}$: (left) as a function of temperature $T$ with chemical potential fixed at $\mu=\{0,\ 200,\ 300\}\ \rm{MeV}$, (right) as a function of chemical potential $\mu$ with temperature fixed at $T=\{50,\ 100,\ 150\}\ \rm{MeV}$. For $g_\omega/m_\omega=0\ \rm{MeV}^{-1}$, the results represent the classical LSM. In case with small coupling constants, the nucleon mass $E_B$ decreases monotonically with increasing temperature and density, similar to the case of the classical model, but this decrease decelerates with the inclusion of vector interactions. Both the classical LSM and cases with small $g_\omega/m_\omega$ experience a rapid decrease in the nucleon mass near $T_c$. For large coupling constants, significant vector interactions lead to a different trend. As demonstrated in~\cite{Jin:2003it}, although in a different system with the degree of freedom being nucleon, the nucleon mass $E_B$ first increases and then decreases with increasing temperature and density, which also reflects a competitive effect between scalar and vector fields. The nucleon mass increases with $g_\omega/m_\omega$, and  the gap between it and the three free quark components decreases, indicating that hadrons become increasingly unstable at high temperatures and densities due to the repulsive force supported by vector interactions.

The mass of hadrons plays an important role in the simulation of relativistic heavy-ion collisions. As an example, the particle yield in the thermal-statistical model $Y \propto \exp(-\frac{\sqrt{m^2+p^2}-\mu}{T})$, which is highly related to hadron mass. The drastic change in hadron mass near the phase boundary may play an important role in accurately explaining and predicting the particle yield during the hadronization process of relativistic heavy-ion collisions. Meanwhile, we can easily get the vector coupling constant by fitting the experimental data with the relationship between the vector coupling constant, hadron mass, and yield.

\section{Summary}

In this work, we investigated the Quark Meson model with vector interactions, incorporating quarks and $\sigma$ mesons, as well as coupling with vector $\omega$ mesons. We solved for the chiral solitons at different temperatures and densities, applying appropriate boundary conditions, and obtained the static properties, including the mass and radius, of nucleons.

The results indicate that the inclusion of vector interactions lead to significant change in the nucleon mass and RMS radius as temperature and density increase. This suggests that the contribution of $\omega$ mesons should be considerable in high temperature and density regions. Specifically, the nucleon mass increases with rising vector coupling constant $g_\omega/m_\omega$, which also reduces the energy gap between the hadronic system and the energy of three free quark components. This indicates that hadrons become increasingly unstable at high temperatures and densities due to vector interactions. Moreover, the repulsive force provided by vector interactions among quarks results in an increase in the root mean square radius of hadrons following the inclusion of $\omega$ mesons. The influence of vector interactions on particle yield in relativistic heavy-ion collisions warrants further investigation, which will be addressed in further research.

\section*{Acknowledgments}

We acknowledge the Guangdong Major Project of Basic and Applied Basic Research (Grant No. 2020B0301030008);
the Science and Technology Program of Guangzhou (Grant No. 2019050001);
the National Natural Science Foundation of China (Grants No. 12105107).

\end{document}